\documentclass{aastex}
\usepackage{spr-astr-addons}
\usepackage{url}
\usepackage{multirow}

\RequirePackage{color}

\newcommand{\emaila}{wangdh@gznu.edu.cn}

\def\pdot{\dot{P}}

\begin{document}

\title{ Evolution implications of neutron star magnetic fields: inferred  from  pulsars and cyclotron lines of HMXBs}
\slugcomment{Not to appear in Nonlearned J., 45.}
\shorttitle{Short article title}
\shortauthors{Autors et al.}

\author{Chang-Qing Ye\altaffilmark{1}} \and \author{De-Hua Wang\altaffilmark{1,*}}\and  \author{Cheng-Min Zhang\altaffilmark{2,3,4}} \and\author{Zhen-Qi Diao\altaffilmark{1}}
\email{\emaila}
\altaffiltext{1}{School of Physics and Electronic Science, Guizhou Normal University, Guiyang 550001, China. $^*$\emaila}
\altaffiltext{2}{CAS Key Lab of FAST, National Astronomical Observatories, Chinese Academy of Sciences, Beijing 100012, China.}
\altaffiltext{3}{School of Physical Sciences, University of Chinese Academy of Sciences, Beijing 100049, China.}
\altaffiltext{4}{Key Laboratory of Radio Astronomy, Chinese Academy of Sciences, Beijing 100012, China.}

\begin{abstract}

The evolution of neutron star (NS) magnetic field (B-field) has long been an important topic, which  is  still  not yet settled down. Here,   we analyze the NS B-fields inferred by the cyclotron resonance scattering features (CRSFs) for the high mass X-ray binaries (HMXBs) and by the magnetic dipole model for the spin-down pulsars.  We find that the B-fields of both the 32 NS-HMXBs and 28 young pulsars with the supernova remnants follow the log-normal   distributions, with
the average values of $3.4\times10^{12}$ G and $4.1\times10^{12}$ G respectively, which are further verified to come from the same continuous distribution by the statistical tests. These results declaim that the two methods of measuring NS B-fields are reliable for the above two   groups of   samples.  In addition, since the NS-HMXBs  have experienced  the spin-down phase as the normal pulsars without accretion  and  then  the  spin-up phase by accretion, their ages should be about million years (Myrs).    Our statistical   facts   imply that  the  B-fields of NS-HMXBs have little decayed in their non-accretion spin-down phases of $\sim$ Myrs, as well as in their accretion phases of $\sim$\,0.1\,Myrs.

\end{abstract}


\section{Introduction}

The magnetic fields (B-fields) of neutron stars (NSs) play the significant roles in many aspects of pulsar phenomena, e.g.,
the electromagnetic emissions from the radio to high energy   wavebands  \citep{har2017,man17,beck2018}.
Until now,   more than  2700 radio pulsars \citep{lor19} and  over  300 accreting NSs in high/low  mass X-ray binaries (HMXBs/LMXBs) \citep{liu2007,wal2015} have been observed.
It is generally accepted that the B-fields of NSs  dramatically decay in the binary accretion phase, while those of the isolated  NSs  should not decay, based on which the low B-field distributions of   the  double NSs ($B \sim 10^9-10^{10}$\,G, see \citealt{yang2019}) and millisecond pulsars (MSPs, $B\sim 10^{7.5}-10^9$\,G, see \citealt{pan18}) are very well understood by the binary accretion \citep{bha91,zhang2016,heu17}. However, there are still debated issues on the evolutions of NS B-fields:  In what extent  the B-fields of  the  isolated NSs have no decay?  If the B-fields decay in the accretion phase, can we estimate the extent of   the  decay for NSs in HMXBs as an inverse relation to the accretion mass proposed \citep{Shi1989,kul1986,zhang06} ? To answer the above questions,  one  could evaluate  the evolution mechanism of NS B-fields, e.g., the Ohmic dissipation \citep{Gep1994}, Hall-drift effect \citep{Gep2002,cummi04}, screening by diamagnetism of accreting matter \citep{cummi2001,Love2005}, and accretion flow to cause the B-field decay \citep{heu95,Mel2001,kon2004,payne04,zhang06}.

  There  exist   the two methods    for   measuring the  NS B-fields, the magnetic dipole model   for the spin-down pulsars \citep{smi08}, and cyclotron resonance scattering features (CRSFs) for NS-HMXBs \citep{mes1992}.   To evaluate the B-field evolution of NSs, we should
  have the good measurements of both the B-fields and their ages, the young pulsars with the  supernova remnants (SNRs) and old NSs in HMXBs. The age of SNR pulsar  is less than 0.3 million years (Myrs) \citep{Fer2012}, which may retain the B-field distribution similar to those as they are born, whereas  the NSs in HMXBs have experienced the spin-down and accretion induced  spin-up phase with the age of $\sim$ Myrs. Thus, the comparison of B-fields of  both groups could acquire  the information of B-field evolutions of NSs.

This paper aims to probe the NS B-field evolution of NS-HMXBs and SNR pulsars   by comparing their B-field distributions, inferred by CRSFs and
 by magnetic dipole model, respectively.
 The structure of the paper is organized as follows:
In Section 2,   we introduce the B-field data  of NS-HMXBs  and  SNR pulsars.
In Section 3,  we analyze the B-field distributions of the two   group  of samples, where the statistical tests are performed to
see their  properties.
Finally, we present the discussions and conclusions in Section 4.

\begin{table*}
\caption{CRSF energy and  inferred B-fields of NS-HMXBs}
\footnotesize
\setlength{\tabcolsep}{1pt}
\begin{tabular}{@{}llccl@{}}
\hline
\noalign{\smallskip}
Source (32) & $E_{\rm c}$ & B-field & $P$ & REFS.$^a$ \\
& (keV) & ($10^{12}$\,G) & (s) & \\
\hline
\noalign{\smallskip}
Swift J1626.6-5156 & 10 & 1.12 & 15.35 & [1, 1] \\\relax
XMMU J05 & 10 & 1.12 & 61.23 & [2, 33] \\\relax
KS 1947+300 & 12.50 & 1.40 & 18.70 & [3, 34] \\\relax
4U 0115+634 & 14, 24, 36, 48, 62 & 1.57 & 3.61 & [4, 35] \\\relax
XTEJ1829-098 & 15 & 1.68 & 7.84 & [5, 5] \\\relax
IGR J17544-2619 & 17 & 1.91 & -- & [6, 36] \\\relax
4U 1907+09 & 19, 39 &  2.12 & 437--440 & [7, 37]\\\relax
4U 1538-52 & 22, 47 & 2.47 & 528--530 & [8, 38] \\\relax
IGR J18179-1621 & 22 & 2.47 & 11.82 & [9, 39] \\\relax
IGRJ18027-2016 & 23 & 2.58 & 139.61  & [10, 40] \\\relax
2S 1553-542 & 23.50 & 2.63 & 9.27 & [11, 11] \\\relax
Vela X-1 & 25, 50 & 2.80 & 283 & [12, 41] \\\relax
V 0332+53 & 27, 51, 74  & 3.03 & 4.37 & [13, 42] \\\relax
SMC X-2 & 27 & 3.03 & 2.37 & [14, 14] \\\relax
Cep X-4 & 28,45 & 3.14 & 66.25 & [15, 43] \\\relax
X per & 29 & 3.25 & 835 & [16, 44] \\\relax
IGR J16393-4643 & 29.3 & 3.28 & 904 & [17, 17] \\\relax
Cen X-3 & 30 & 3.36 & 4.82 & [18, 45] \\\relax
J 16493-4348 & 30 & 3.36 & 1093 & [19, 46] \\\relax
RX J0520.5-6932 & 31.50 & 3.53 & 8.03 & [20, 20] \\\relax
RX J0440.9-4431 & 32 & 3.59 & 202.50 & [21, 47] \\\relax
MXB 0656-072 & 33 & 3.70 & 160 & [22, 48] \\\relax
IGR J19294+1816 & 35.50 & 3.98 & 12.44 & [23, 49] \\\relax
XTE JI1946+274 & 36 & 4.03 & 15.80 & [24, 50] \\\relax
4U 1626-67 & 37 & 4.15 & 7.67 & [25, 51] \\\relax
GX 301-2 & 37 & 4.15 & 675--700 & [26, 52] \\\relax
Her X-1 & 39, 73 & 4.37 & 1.24 & [27, 53] \\\relax
MAXI J1409-619 & 44, 73, 128  & 4.93 & 500 & [28, 54] \\\relax
1A 0535+262 & 45, 100 & 5.04 & 103 & [29, 55] \\\relax
GX 304-1 & 54 & 6.05 & 272 & [30, 56] \\\relax
1A 1118-615 & 55, 112? &  6.16 & 405--407 & [31, 57] \\\relax
GRO J1008-57 & 76 & 8.52 & 93.60 & [32, 58] \\
\hline
\multicolumn{5}{l}{
Note: $E_{\rm c}$---CRSF energy of the fundamental and multiple} \\
\multicolumn{5}{l}{harmonics; $P$---Spin period of NS; $^a$---The first reference of} \\
\multicolumn{5}{l}{each source is for $E_{\rm c}$ (for some sources it record the} \\
\multicolumn{5}{l}{fundamental and multiple harmonics CRSFs observed} \\
\multicolumn{5}{l}{simultaneously), and the second reference is for $P$.} \\
\end{tabular}
\begin{tabular}{@{}l@{}}
\begin{minipage}{80mm}
REFS.:
[1]. \citealt{dec2013};
[2]. \citealt{mano2009a};
[3]. \citealt{furs2014a};
[4]. \citealt{whea1979a,hein1999a,ferr2011a};
[5]. \citealt{sht2019};
[6]. \citealt{bhal2015a};
[7]. \citealt{rive2010a,cusu1998a};
[8]. \citealt{rod2009,clar1990a};
[9]. \citealt{lij2012a,Tuer2012a};
[10]. \citealt{luto2017a};
[11]. \citealt{tsyg2016a};
[12]. \citealt{laba2003a,krey1999a,kend1992a};
[13]. \citealt{naka2010a,maki1990a,cobu2005a};
[14]. \citealt{jais2016a};
[15]. \citealt{jai2015,miha1991a};
[16]. \citealt{cobu2001a};
[17]. \citealt{boda2016a};
[18]. \citealt{burd2000a,naga1992a};
[19]. \citealt{daia2011a};
[20]. \citealt{tend2014a};
[21]. \citealt{tsyg2012a};
[22]. \citealt{mcbr2006a,hein2003a};
[23]. \citealt{jay2017};
[24]. \citealt{hein2001a};
[25]. \citealt{orla1998a};
[26]. \citealt{suc2012,maki1992a};
[27]. \citealt{enot2008a,tru1978};
[28]. \citealt{orl2012};
[29]. \citealt{tera2006a};
[30]. \citealt{yama2011a,miha2010a};
[31]. \citealt{doro2010a};
[32]. \citealt{yama2014a};
[33]. \citealt{man2009};
[34]. \citealt{chak1995,gall2004};
[35]. \citealt{rapp1978}
[36]. \citealt{clar2009}
[37]. \citealt{intz1998}
[38]. \citealt{clar2000,clar1994}
[39]. \citealt{bozz2012};
[40]. \citealt{maso2011};
[41]. \citealt{quai2003};
[42]. \citealt{zhangs2005,qjl2005};
[43]. \citealt{miha1991a};
[44]. \citealt{whit1976,delg2001};
[45]. \citealt{meer2007};
[46]. \citealt{pear2013};
[47]. \citealt{liu2006};
[48]. \citealt{mcbr2006a};
[49]. \citealt{rodr2009,corb2009};
[50]. \citealt{smit1998,wils2003};
[51]. \citealt{tsu1986};
[52]. \citealt{kape2006,kohd1997};
[53]. \citealt{trum1978};
[54]. \citealt{yam2010};
[55]. \citealt{rose1975,fing1996};
[56]. \citealt{huck1977,peri1983};
[57]. \citealt{stau2011};
[58]. \citealt{stol1993,kueh2012}.
\end{minipage}
\end{tabular}
\label{tab1}
\end{table*}

\begin{table*}
\caption{B-field statistics of 32 NS-HMXBs and 28 SNR pulsars}
\resizebox{110mm}{!}{
\begin{tabular}{llcccccc}
\hline
Category & Number & Range of B & $\langle B\rangle$ & $\tilde{B}$ & $\sigma_{B}$ \\
\cline{3-6} \\
 & & & ($\times10^{12}$\,G) \\
\hline
 {\rm NS-HMXBs}   & 32 & $1.12\sim8.52$ & 3.39 & 3.27 & 1.58 \\
\hline
{\rm SNR  pulsars} & 28 & $1.11\sim9.91$ & 4.09 & 3.60 & 2.14 \\
\hline
\end{tabular}}\\
Note:  $\langle B\rangle$---mean of $B$;  $\tilde{B}$---median of $B$;  $\sigma_{B}$---standard deviation of $B$.
\label{tab2}
\end{table*}

\section{Two methods of NS magnetic field measures}

\begin{figure}[t]
\includegraphics[width=8.3cm]{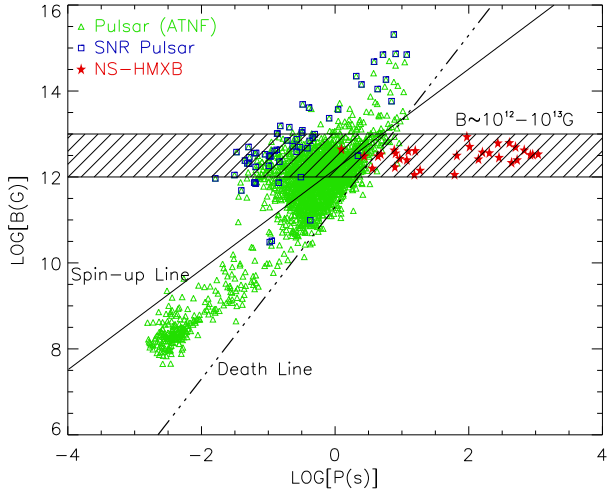}
\caption{ Diagram  of the NS B-field versus spin period.
The data of NS-HMXBs and pulsars are taken from Table \ref{tab1} and ATNF pulsar catalog \citep{man05}, respectively.
The shadow area indicates the $B$ range of $\sim10^{12}-10^{13}$\,G, in which the B-fields of NS-HMXBs are measured.
The solid line and   dashed-dot line represent the spin-up line and death line of pulsars  as defined by \citet{bha91}.}
\label{fig1}
\end{figure}

\subsection{B-fields of NS-HMXBs by CRSFs}
For the accretion powered X-ray pulsars in HMXBs, NSs emit X-rays (e.g., $\sim1-100$\,keV)   by accreting matters from their  companions \citep{cab2012, wal2015}. In the strong B-field regime of NS surface, the electron energy is quantized to the discrete Landau levels,   which experiences the resonance scattering with photons and  results in the absorption features in the X-ray energy spectrum \citep{tru1978},   i.e., the energy  of CRSFs:
\begin{equation}
E_{\rm c}=\hbar \frac{eB}{m_{e}c}=11. 6(\rm keV)\times B_{12} 
\label{eq1}
\end{equation}
where $B_{12}\equiv \frac{B}{10^{12}\,\rm G}$, with the electron mass $m_e$ and charge $e$, as well as the speed of  light $c$ and Planck constant $\hbar$.
The CRSFs are  generally  considered to form  around  NS surface \citep{mes1992, sch2007, bec2012},
so the observed energy $E_{\rm c}$ are gravitationally red-shifted and corrected to
\begin{equation}
E_n=nE_{\rm c}=(1+z)E_{n, \rm obs}
\label{eq2}
\end{equation}
where $n$ is the quantum number of   the  energy level, corresponding to the  fundamental  and harmonics of  CRSF  given by $n=1,2,3,\ldots$, and the parameter  $z$ is the gravitational red-shift expressed as,
\begin{equation}
z=\frac{1}{\sqrt{1-\frac{2GM}{Rc^{2}}}}-1\;,
\label{eq3}
\end{equation}
where $G$ is the Newtonian gravitational constant. For the typical NS mass of $M=1.4\,{\rm M_\odot}$ and radius of $R=10$\,km, the gravitational red-shift parameter is obtained as $z=0.3$ \citep{sha83,mai2017}.

We search for the NS-HMXBs which have been detected with CRSFs, and focus on the 32 sources with the confirmed detection.
  Table \ref{tab1} lists the CRSF energies ($E_{\rm c}$, fundamental and harmonic) and NS spin periods ($P$) of these sources that are collected from the published literature \citep{mak1999, Coburn2002, cab2012, wal2015}.
Furthermore,  we infer the B-fields of  NS-HMXBs by equation (\ref{eq1})-(\ref{eq3}) with the CRSFs in Table \ref{tab1}  by assuming they emit around the NS surface \citep{mes1992,bec2012}.  Figure \ref{fig1} illustrates the NS B-field versus spin period distribution of the samples.

\subsection{B-fields of    pulsars by their periods and derivatives}
For the spin-down  pulsars, their  losses of rotational  energies  is presumed  by the  magnetic diploe radiation, based on which their  B-fields   can be derived by  their observed spin  periods ($P$) and derivatives $\pdot$,  as given in the references in the case of perpendicular rotator \citep{ost1969, man1977, smi08}:

\begin{equation}
B=(\frac{3I\rm c^{3}P\dot{P}}{8{\rm \pi^{2}}R^{6}})^{1/2}
\simeq 3.2\times10^{19}(P\dot{P})^{1/2}\;({\rm G}),
\label{eq4}
\end{equation}
where $I$ and $R$ stand for the moment of inertia and radius of NS with the  conventional values of
 $I= 10^{45}\rm g\cdot cm^{2}$ and $R=10$\,km.

In order to investigate the   B-field  evolution of NSs, we   compare the B-field distribution of the NS-HMXBs with
that of the young pulsars with   SNRs,   whose ages  are generally less than 0.3\,Myr. 
Until now,  there are  61 SNR pulsars have been   recorded  in ATNF pulsar catalog \citep{man05},  in which 28 sources  share the B-fields in the
 ranges of $\sim10^{12}-10^{13}$\,G (as compared to those of CRSF sources, see Table 1) that are analyzed in the paper.   These samples are
    labeled in the diagram of  B-field versus spin period, as  plotted in Figure \ref{fig1}.

\begin{table}[htp]
\caption{S-W Test  Results of  NS B-field Distributions}
\begin{center}
\small
\setlength{\tabcolsep}{2pt}
\resizebox{80mm}{!}{
\begin{tabular}{lcccc@{}}
\hline
Category & Number & $S-W$& Reject $H_{0}$ \\
 & & ($p$-value) \\
\hline
NS-HMXB & 32 &   0.56  & No \\
\hline
\noalign{\smallskip}
SNR pulsar & 28 & 0.97 & No \\
\hline
\end{tabular}}
\begin{minipage}{80mm}
Note:
  $H_0$ is the null hypothesis that the data follows a log-normal
distribution, with the confidence level parameter $\alpha=0.05$.
\end{minipage}
\label{tab3}

\end{center}
\end{table}

\begin{table}[htp]
\caption{K-S Test Results of NS B-field Distributions}
\small
\setlength{\tabcolsep}{2pt}
\resizebox{80mm}{!}{
\begin{tabular}{@{}lccc@{}}
\hline
\noalign{\smallskip}
Category & Number & $K-S$ & Reject $H_{0}$ \\
 & & ($p$-value) & \\
\hline
\noalign{\smallskip}
NS-HMXB & 32 & \multirow{2}*{$0.74$} & \multirow{2}*{No} \\
SNR pulsar & 28 & & \\
\hline
\end{tabular}}
\begin{minipage}{80mm}
Note:
$H_0$ is the null hypothesis that the two groups of data come from the same
continuous distribution, with the confidence level parameter $\alpha=0.05$.
\end{minipage}
\label{tab4}
\end{table}

\section{Analysis on the B-field distributions}

 The  NS B-field statistical quantities  of the samples , including the range, mean ($\langle B \rangle$), median ($\tilde{B}$) and standard deviation ($\sigma_{B}$), are obtained and  summarized in Table \ref{tab2}.
It can be seen that the NS-HMXBs share the similar $\langle B \rangle$  and  $\tilde{B}$  values  ($\langle B \rangle\sim3.39\times10^{12}$\,G  and $\tilde{B}\sim3.27\times10^{12}$\,G)  to those of the SNR pulsars  ($\langle B \rangle\sim4.09\times10^{12}$\,G  and  $\tilde{B}\sim3.60\times10^{12}$\,G). In order to investigate  whether the two group samples of B-fields come from the same continuous distribution, we make the following two steps: (I).  testing the two distributions to be the log-normal or not, and (II). testing the two group B-fields to come from the same continuous distribution or not if both distributions follow the log-normal.

 (I). Table \ref{tab3} lists the Shapiro-Wilk (S-W) test results \citep{cord2009} of  the two group B-field samples,  from which it
    can be noticed  that the B-fields of both   the  NS-HMXBs and SNR pulsars  are compatible with the log-normal distributions at the 95 percent confidence level. The expressions of the mean and standard deviation of the log-normal distributions are $(28.75\pm0.47)\,{\rm log}_e{\rm (G)}$ for the NS-HMXBs, and ($28.91\pm0.52)\,{\rm log}_e{\rm (G)}$ for the SNR pulsars, respectively.

\begin{figure}[t]
\includegraphics[width=8.3cm]{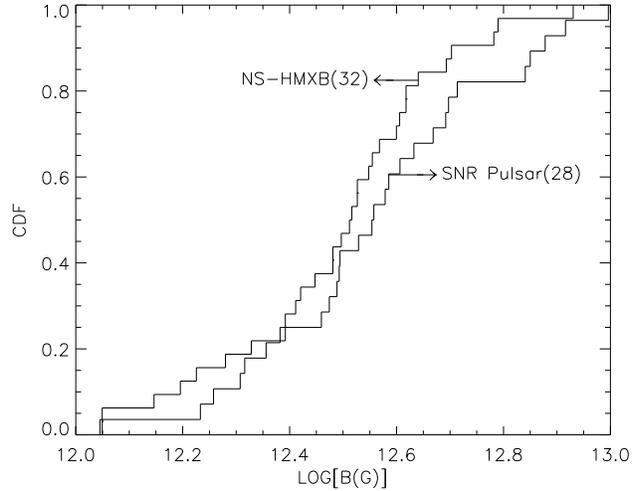}
\caption{The CDF curves  of B-fields  for the NS-HMXBs and SNR pulsars. }
\label{fig2}
\end{figure}

 (II). Furthermore, we take the Kolmogorov-Smirnov (K-S) test \citep{cord2009} to compare the B-field distributions of the NS-HMXBs and SNR pulsars,  and  show the results in Table \ref{tab4}, where the consistency of both distributions is obtained at the 95 percent confidence level. In  order to more intuitively represent the two group B-fields come from  the same log-normal distribution, we also show the corresponding B-field cumulative distribution function (CDF) curves of the two group samples in Figure \ref{fig2}, the appearance of which clearly strengthens the conclusion of the same distributions.

\section{Discussions and Conclusions}

 In order to  investigate whether the NS B-field decay or evolute in million years, we studied the two distinctive  classes of NS samples:
NS-HMXBs that are the accretion X-ray binary systems with the ages of $\sim$ Myrs \citep{bha91}, and SNR pulsars that are the
pulsating objects with the ages of less than 0.3\,Myrs \citep{Fer2012}.  The B-fields of  the formers are measured directly by CRSFs and
those of the latters  are derived by the magnetic dipole model. The following facts and conclusions are noticed and obtained:

\begin{itemize}
\item
  The S-W test indicates that the  B-fields of both the  32 NS-HMXBs and 28 SNR pulsars follow the log-normal distributions (see Table \ref{tab3}), and the two group B-fields  are further verified to come from the same  continuous distribution by the K-S test (see Table \ref{tab4}).
  According to the evolutional theory of the X-ray binaries,  the NSs in HMXBs may firstly experience the spin-down phase   for  $\sim$ Myrs by the magnetic dipole emissions \citep{bha91,lor08}, and  then they can switch to the accretion induced spin-up phase for $\sim$ 0.1\,Myr \citep{bha91}.
  Therefore, the statistical test  facts imply that  the B-fields of NSs in HMXBs have little  decayed in their non-accretion spin-down phases of $\sim$\,Myrs,  as well as in their accretion phases of $\sim$\,0.1\,Myrs.
  Furthermore, it is true that the NS B-field decays in the binary  accretion  phase present a positive correlation with how much mass accreted, where the surface field decreases both by Ohmic decay and submergence by the inflowing matter (\citealt{bisno74,Shi1989,bha91}; \citealt{Gep1994,Urpin95}; \citealt{zhang06}). In particular, it is suggested that for an accretion rate of $\sim10^{-10}\,{\rm M_\odot\,yr^{-1}}$, which corresponds to an accreted mass of $\sim10^{-5}\,{\rm M_\odot}$ during $\sim0.1$\,Myr, the NS surface field may decay by $\sim$ two orders of magnitude \citep{Urpin95,Urpin98}. Therefore, the consistent B-field distributions between the SNR pulsars and NS-HMXBs imply that the actually accreted mass of HMXBs may be less than $\sim10^{-5}\,{\rm M_\odot}$.

\item It should be noticed that the B-field inferred from the CRSF measurement is the B-field around the NS surface, while the B-field obtained from the measurement of $P$ and $\dot{P}$ values is calculated by considering the energy loss at the NS light cylinder and then extrapolates the field strength back to the star's surface assuming the flat space-time $1/r^3$  radial dependence of the field. Thus, for the HMXBs, equation (\ref{eq2}) takes into account gravitational red-shift, while general relativity effects are not taken into consideration for the B-field of isolated pulsars. The different B-field measures for the NS-HMXBs and SNR pulsars present the same distributions,
  and we declaim that the two methods, by the CRSF and  magnetic  dipole model, are successfully applied to obtain the B-fields around  NS surfaces.
  In general, the CRSF  indicates the local B-field around  NS surface, while the B-field inferred  by the magnetic dipole
  model represents the large scale value.  The similarity  of both B-field values might imply that both  the measures and selected NS
  parameters should be in the reasonable regimes, at least valid in the statistical sense.

 \end{itemize}
   The  B-field evolution mechanism of NS is critical for understanding the   emissions  and binary interactions of various pulsars, thus
   our  conclusions, based on the B-field measures of different methods and statistical tests,
     would  provide the robust  clues  or constraints for the time scale  of B-field decay.

\acknowledgments
This work is supported by the National Natural Science Foundation of China (Grant No. 11703003 and No. U1731238), the Science and Technology Foundation of Guizhou Province (Grant No. [2017]5726), NAOC-Y834081V01.

\end{document}